\documentclass[]{mn2e}

\usepackage{amssymb}
\usepackage{amsmath}
\usepackage{graphicx}
\newcommand{\mnras}{MNRAS}
\newcommand{\apj}{ApJ}

\usepackage{appendix}
\usepackage{longtable}
\usepackage{url}

\title[LF dependence on profile]{The massive end of the luminosity and stellar mass functions:  Dependence on the fit to the light profile}
\author[Bernardi et al.]{\parbox{\textwidth}{M. Bernardi$^{1}$\thanks{E-mail: bernardm@sas.upenn.edu},
A. Meert$^{1}$, R. K. Sheth$^{1,2}$, V. Vikram$^{1}$, M. Huertas-Company$^{3}$, 
S. Mei$^{3}$ \& F. Shankar$^{3}$}\vspace{0.4cm}\\
\parbox{\textwidth}{$^{1}$Department of Physics and Astronomy, University of Pennsylvania, 
Philadelphia, PA 19104, USA\\
$^{2}$The Abdus Salam International Center for Theoretical Physics, 
      Strada Costiera 11, 34151 Trieste, Italy\\
$^{3}$GEPI, Observatoire de Paris, CNRS, Univ. Paris Diderot;
Place Jules Janssen, 92190 Meudon, France\\ }}

\begin{document}
 \date{Accepted .  Received ; in original form }

\maketitle

\label{firstpage}

\begin{abstract}
In addition to the large systematic differences arising from assumptions about the stellar mass-to-light ratio, the massive end of the stellar mass function is rather sensitive to how one fits the light profiles of the most luminous galaxies.  We quantify this by comparing the luminosity and stellar mass functions based on SDSS {\tt cmodel} magnitudes, and {\tt PyMorph} single-Sersic and Sersic-Exponential fits to the surface brightness profiles of galaxies in the SDSS.  The {\tt PyMorph} fits return more light, so that the predicted masses are larger than when {\tt cmodel} magnitudes are used.  As a result, the total stellar mass density at $z\sim 0.1$ is about $1.2\times$ larger than in our previous analysis of the SDSS.  The differences are most pronounced at the massive end, where the measured number density of objects having $M_*\ge 6\times 10^{11}M_\odot$ is $\sim 5\times$ larger.  Alternatively, at number densities of $10^{-6}$Mpc$^{-3}$, the limiting stellar mass is $2\times$ larger.  The differences with respect to fits by other authors, typically based on Petrosian-like magnitudes, are even more dramatic, although some of these differences are due to sky-subtraction problems, and are sometimes masked by large differences in the assumed $M_*/L$ (even after scaling to the same IMF).  Our results impact studies of the growth and assembly of stellar mass in galaxies, and of the relation between stellar and halo mass, so we provide simple analytic fits to these new luminosity and stellar mass functions and quantify how they depend on morphology, as well as the binned counts in electronic format.  While these allow one to quantify the differences which arise because of the assumed light profile, and we believe our Sersic-Exponential based results to be the most realistic of the models we have tested, we caution that which profile is the most appropriate at the high mass end is still debated.  

\end{abstract}

\begin{keywords}
 galaxies: fundamental parameters -- galaxies: luminosity function, mass function -- galaxies: photometry
\end{keywords}

\section{Introduction}
The brightest, most massive galaxies have been the object of much study.  Recent work has emphasized the importance of using a good parametrization of the abundance at the bright, massive end if one is interested in using Halo Model based abundance matching techniques, or extreme value statistics, to understand their origin (e.g. Paranjape \& Sheth 2012).  A few years ago Bernardi et al. (2010) noted that the most luminous galaxies were more abundant than expected from the most commonly used parametrizations of the luminosity function.  They also pointed out that, when converted to a stellar mass function, this mis-match was important for models which use the observed abundance and its evolution to constrain the issue of whether these objects were assembled via major or minor mergers.  However, they also showed that the conversion from $\phi(L)$ to $\phi(M_*)$ is rather sensitive to the assumed stellar mass-to-light ratio, for which, as we show below, there is still no consensus.  

Bernardi et al. (2010) used luminosities estimated from the {\tt cmodel} magnitudes output by the Sloan Digital Sky Survey (hereafter SDSS, Abazajian et al. 2009).  These tended to return more light than the more commonly used estimates based on the Petrosian radius defined by the SDSS, especially for the brightest objects, although some of this difference was due to sky subtraction problems in the SDSS.  Bernardi et al. (2010) applied a crude correction for this to the {\tt cmodel} magnitudes, but not to the Petrosian magnitudes output by the SDSS pipelines, primarily because essentially all previous work with Petrosian magnitudes made no such correction.  

The {\tt cmodel} magnitudes are a poor-man's best guesstimate for the total light if the surface brightness distribution of the objects follows neither a pure exponential disk nor a deVaucouleur's profile (Bernardi et al. 2007).  Recently, Meert et al. (2013a,b) have performed more careful Sersic-bulge + exponential disk (+ sky) decompositions of these objects.   These typically return even more light than the {\tt cmodel} magnitudes (e.g. Bernardi et al. 2013), in part because of the improved treatment of the sky, but also because differences in the model which is fitted to the observed light profile matter.  

The main purpose of the present note is to show how these differences impact estimates of the luminosity and stellar mass functions at the bright end.  As one might expect, the effect is at least as dramatic as the choice of $M_*/L$.  Therefore, a related goal of the present work is to separate out the effect on $\phi(M_*)$ of how the luminosity was estimated from that of $M_*/L$.  

Section~\ref{lfmf} describes our sample, shows the luminosity and stellar mass functions, quantifies how they depend on the fit to the light profile and provides simple fitting formulae which quantify our results as well as the binned counts in electronic format.  While these results allow one to easily account for the dependence on the light profile (e.g. using Sersic instead of SDSS {\tt cmodel} or Petrosian magnitudes), the question of which $M_*/L$ estimate is most appropriate is beyond the scope of this work, and deserves further study.  For reasons described in Bernardi et al (2010), all our $M_*/L$ estimates assume a Chabrier (2003) IMF.  In Section~\ref{comparison} we show that, even though a number of recent works have made this same choice for the IMF (Baldry et al. 2012; Moustakas et al. 2013), they still have $M_*/L$ values which are very different from ours (i.e., Bernardi et al. 2010), from one another, and from earlier work (Bell et al. 2003).  That is to say, differences in $M_*/L$ arise even when the same IMF is assumed:  this is not generally appreciated.  In Section~\ref{morph} we show how the luminosity and stellar mass functions depend on morphological type, where the type is determined by the Bayesian Automated Classification scheme of Huertas-Company et al. (2011).  A final section summarizes.   

\begin{figure}[b]
 \centering
 \includegraphics[scale = .4]{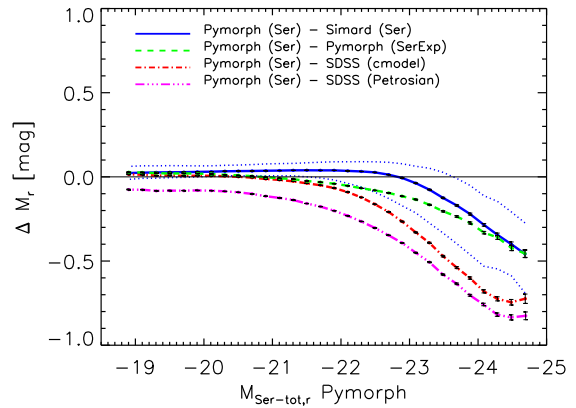}
 \caption{Difference between {\tt PyMorph} Sersic fits and SDSS DR7 Petrosian, SDSS {\tt cmodel}, {\tt PyMorph} SerExp, and Sersic fits from Simard et al. (2011) (bottom to top), for galaxies in the sample selected by Bernardi et al. (2010).  Petrosian magnitudes are always the faintest, whereas single Sersic-based magnitudes tend to be the brightest. Dotted lines around {\tt PyMorph} (Ser)--Simard (Ser) show the 16th and 84th percentiles of the distribution; these are 
similar to the scatter around the median for the other curves. }
 \label{compfit}
\end{figure}

When converting from apparent brightnesses to luminosities, we assume a spatially flat background cosmology dominated by a cosmological constant, with parameters $(\Omega_m,\Omega_\Lambda)=(0.3,0.7)$, and a Hubble constant at the present time of $H_0=70$~km~s$^{-1}$Mpc$^{-1}$.

\section{Luminosity and stellar mass functions}\label{lfmf}

\subsection{The sample}
To provide a direct comparison with previous work, we have selected the same sample as Bernardi et al. (2010); i.e., about 260,000 SDSS galaxies having $14.5\le m_{r{\rm Pet}}\le 17.7$.  We obtained the Petrosian and {\tt cmodel} estimates of the total light for each of these objects from the SDSS DR7 database.  These are known to suffer from sky-subtraction and crowded-field/masking problems (Bernardi et al. 2010; Meert et al. 2013a,b).  In what follows, the {\tt cmodel} magnitudes we use are crudely corrected for the SDSS sky subtraction problems as described in Bernardi et al. (2010).  On the other hand, analogous corrections to the Petrosian magnitudes are rarely made, so, for ease of comparison with previous work, we apply no such correction here (we discuss this further in the context of Figure~\ref{compfit}).  

We then ran {\tt PyMorph} (Vikram et al. 2010; Meert et al. 2013a) on these objects.  This is an algorithm which uses {\tt GALFIT} (Peng et al. 2002) to fit seeing-convolved 2-dimensional Sersic + exponential models to the observed surface brightness profiles of galaxy images.  Results from extensive tests indicate that the algorithm is accurate (Meert et al. 2013a); it does not suffer from the sky-subtraction problems which plague the simpler SDSS reductions especially in crowded fields.  {\tt PyMorph} sometimes fails to converge to an answer; this happens about 2\% of the time, but because this fraction is independent of magnitude, it does not affect our completeness, other than by a small overall scaling.  Finally, we computed $k$- and evolution corrections for each object following Bernardi et al. (2010), and hence, luminosities.  

\begin{figure*}
 \centering
 \includegraphics[scale = .6]{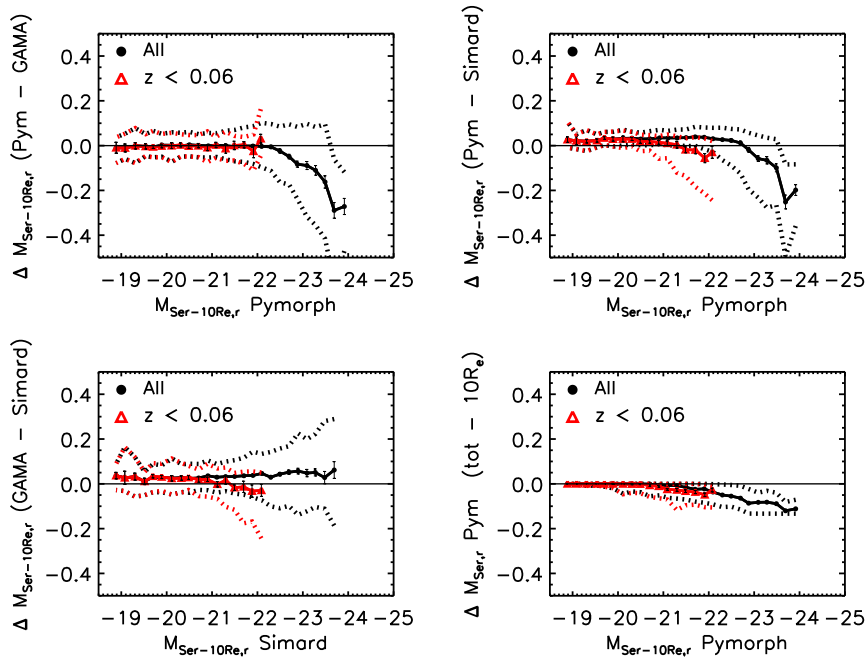}
 \caption{Comparison of single Sersic reductions for the SDSS galaxies in 
          common to Simard et al. (2011), Kelvin et al. (2012) and 
          {\tt PyMorph} (our notation $M_{Ser-10R_e}$ reflects the fact 
          that Kelvin et al. truncate the profile at 10$R_e$).  
          Red triangles show a similar analysis if one restricts to 
          objects with $z<0.06$ as done in Baldry et al. (2012); 
          this shallower volume does not probe the highest luminosities 
          that are of most interest here. Dotted lines show the 16th and 84th 
          percentiles of the distribution.}
 \label{comp}
\end{figure*}

\subsection{Dependence on assumed surface brightness profile}
The magnitudes and half-light radii output by {\tt PyMorph} depend on the model which is fit.  E.g., fitting what is really a two-component image with a single deVaucouleurs profile will generally underestimate the total light.  On the other hand, the total light associated with the best-fit single Sersic or a two-component Sersic-bulge + exponential-disk model, is less-biased from its true value (e.g. Bernardi et al. 2007; Bernardi et al. 2013).  Meert et al. (2013a) and Bernardi et al. (2013) have also shown that in objects brighter than $L_*$, fitting a two-component Sersic  + exponential model to what is really just a single Sersic results in a noisier recovery of the input parameters, but these are not biased.  On the other hand, fitting a single Sersic to what is truly a two-component system results in significant biases.

Although the Sersic + Exponential model is more accurate, the Sersic fit is often performed on real data when it is believed that the resolution and S/N are such that it is unlikely to recover a robust two-component fit. Therefore, since either of these models are expected to be more realistic than a single deVaucouleurs model, we will use both in what follows.

The estimates of the total light from a Sersic or Sersic + exponential model are generally larger than those based on the {\tt cmodel} magnitudes output by the SDSS pipelines (e.g. Bernardi et al. 2007; Hill et al. 2011; Bernardi et al. 2013; also see Mosleh, Williams \& Franx 2013), and both are larger than the SDSS DR7 Petrosian magnitudes.  Figure~\ref{compfit} illustrates that this difference can be large.  Some of this is due to the difference in the treatment of the sky, and some to the differences between the fitted models.  For example, the offset at the faint end between the Petrosian and the other models is almost entirely due to the fact that the SDSS DR7 pipeline tended to overestimate the contribution from the sky, thus making the SDSS Petrosian magnitudes about 0.05~mags too faint.  

After we had completed our study, He et al. (2013) quantified the effects of sky-subtraction and masking problems on the SDSS DR7 Petrosian values:  accounting for these makes their Petrosian magnitudes 0.05~mags brighter at the faint end, and 0.2~mags brighter at the bright end.  (Our own reanalysis, based on {\tt PyMorph} sky-estimates suggests this difference is slightly smaller:  about 0.1~mags at $M_r\le -23.5$.)  As a result, at the bright end, He et al's Petrosian magnitudes are slightly brighter than our {\tt cmodel} magnitudes (recall that the {\tt cmodel} magnitudes include only a crude correction for the SDSS sky subtraction problems), but they are generally fainter than our Sersic or SerExp values at the bright end.  Fundamentally, this can be traced to the well-known facts that (a) Petrosian magnitudes underestimate the total light when the light profile has extended wings, and (b) this is particularly an issue at the bright end (e.g. Binggeli \& Cameron 1991; Blanton et al. 2001; Trujillo et al. 2001; Andreon 2002; Brown et al. 2003; Graham et al. 2005).  At the bright end, this leads to an underestimate of order $\sim 0.3$ mags or more (i.e., this matters more than the sky-subtraction problems), which is similar to the difference between the {\tt cmodel} and SerExp magnitudes.  This expected difference is consistent with He et al.'s finding that even their revised Petrosian magnitudes are systematically fainter than aperture fluxes based on deeper photometry which reaches to 1\% of the sky.


Although the dependence on the assumed light profile is what has motivated our study, it is reasonable to ask if these differences are indeed larger than those associated with different pipelines which fit the same model.  We address this in the next subsection.  

\subsection{Dependence on pipeline}
As a check of our reductions, we have also used luminosities from the single-Sersic-based photometric reductions of Simard et al. (2011).  Figure~\ref{compfit} shows that these are in good agreement with {\tt PyMorph} except for a small offset ($\sim 0.05$~mags), although the differences become large at the bright end.  See Figures~A1 and A2 in Bernardi et al. (2013) and discussion on sky estimates in Meert et al. (2013a,b) for why we believe our estimates are less biased.  In any case, these differences are small compared to {\tt PyMorph}--{\tt cmodel}.  

\begin{table*}
\caption[]{Parameters of $\phi(L_r)$ (top rows) and $\phi(M_*)$ (bottom rows)
           derived from fitting equations~(\ref{phiX}) to the observed 
           counts based on different magnitudes.\\}
\begin{tabular}{lcccccccc}
 \hline 
  Fit & $\phi_*$ & $L_*$ & $\alpha$ & $\beta$ & $\phi_{\gamma}$ & $L_{\gamma}$ & $\gamma$ &  $\rho_L$ \\ 
 & $10^{-2}{\rm Mpc}^{-3}$ & $10^9\,L_\odot$ & & & $10^{-2}{\rm Mpc}^{-3}$ & $10^9\,L_\odot$ & & $10^9\,L_\odot$Mpc$^{-3}$\\
 \hline
cmodel & $ 0.928$ & $   0.3077$ & $   1.918$ & $ 0.433$ & $ 0.964$ & $   1.8763$ & $   0.470$ & $   0.136$ \\
Sersic & $ 1.343$ & $   0.0187$ & $   1.678$ & $ 0.300$ & $ 0.843$ & $   0.8722$ & $   1.058$ & $   0.150$ \\
SerExp & $ 1.348$ & $   0.3223$ & $   1.297$ & $ 0.398$ & $ 0.820$ & $   0.9081$ & $   1.131$ & $   0.146$ \\
Sersic (Simard) & $ 1.920$ & $   6.2456$ & $   0.497$ & $ 0.589$ & $ 0.530$ & $   0.8263$ & $   1.260$ & $   0.152$ \\
 \hline &&&&\\
  Fit & $\phi_*$ & $M_*$ & $\alpha$ & $\beta$ & $\phi_\gamma$ &  $M_\gamma$ & $\gamma$ & $\rho_{M_s}$\\ 
  & $10^{-2}{\rm Mpc}^{-3}$ & $10^9\,M_\odot$ & & & $10^{-2}{\rm Mpc}^{-3}$ &$10^9\,M_\odot$ & & $10^9\,M_\odot$Mpc$^{-3}$\\
 \hline
cmodel & $ 0.766$ & $   0.4103$ & $   1.764$ & $   0.384$ & $ 0.557$ & $   4.7802$ & $0.053$ & $   0.276$ \\
Sersic & $ 1.040$ & $   0.0094$ & $   1.665$ & $   0.255$ & $ 0.675$ & $   2.7031$ & $0.296$ & $   0.344$ \\
SerExp & $ 0.892$ & $   0.0014$ & $   2.330$ & $   0.239$ & $ 0.738$ & $   3.2324$ & $0.305$ & $   0.330$ \\
Sersic (Simard) & $ 0.820$ & $   0.0847$ & $   1.755$ & $   0.310$ & $ 0.539$ & $   5.2204$ & $0.072$ & $   0.349$ \\
 \hline &&&&
\end{tabular}
\label{fits} 
\end{table*}

The {\tt PyMorph} and Simard et al. luminosities come from integrating the fitted profile to infinity.  Other authors truncate, typically at some multiple of the half-light radius.  For example, the analysis of galaxies in the GAMA survey (Galaxy And Mass Assembly survey -- Kelvin et al. 2012) truncates the fits at $10R_e$.  Using the GAMA DR1 data release (Driver et al. 2011), we compare {\tt PyMorph}, Simard, and GAMA values for the ∼ 7335 galaxies for which all three reductions are available. (This sample is set by the fact that the GAMA DR1 covers 100 sq.deg. of the SDSS. GAMA has 10750 matches with the DR7 SDSS spectroscopic galaxy sample, of which 7335 galaxies are in the Bernardi et al. 2010 sample we study here.) Figure~\ref{comp} compares {\tt PyMorph}, Simard, and GAMA values for the single-Sersic magnitude. The bottom right panel shows that the truncation matters at the level of 0.05~mags only at $M_r<-22$.  But otherwise, if truncated similarly, then GAMA and {\tt PyMorph} are in good agreement at $M_r>-22$, whereas {\tt PyMorph} returns significantly more light than the other two at the bright end.  We believe the differences at the bright end are similar in origin (i.e. sky subtraction issues) to those with respect to Simard et al. (see Meert et al. 2013a,b; Bernardi et al. 2013). 


\subsection{The luminosity function}
For each of the estimates of the total light shown in Figure~\ref{compfit}, we estimated the luminosity function as in Bernardi et al. (2010) using the $V_{\rm max}$ method of Schmidt (1968).  (I.e., we weighted each galaxy using $1/V_{\rm max}(L_{r{\rm Pet}})$, where $V_{\rm max}$ is the maximum comoving volume within which the object could have been included in the sample, accounting for both the bright and faint magnitude limits.)  

\begin{figure}
 \centering
 \includegraphics[scale = .4]{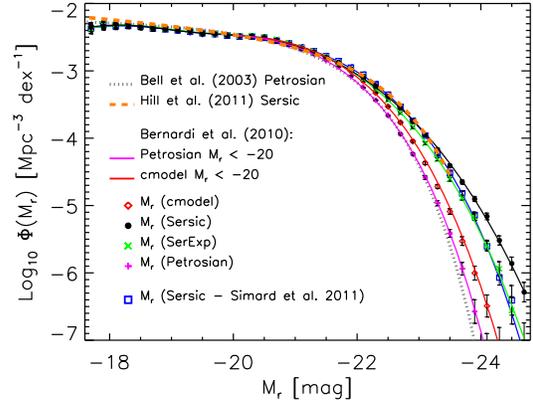}
 \caption{SDSS Main galaxy luminosity function based on  
 Petrosian, {\tt cmodel}, single Sersic from Simard et al. (2011) 
 and {\tt PyMorph} SerExp and Sersic magnitudes   
 (bottom to top at $M_r=-24$).  Smooth curves show the result of 
 fitting equation~(\ref{phiX}) to the counts; associated best-fit 
 parameter values are given in Table~\ref{fits}.  For the Petrosian
 and {\tt cmodel} magnitudes, the curve shown is that reported by 
 Bernardi et al. (2010) on the basis of fitting to $M_r<-20$.
 The Petrosian and Sersic based fits of Bell et al. (2003) and 
 Hill et al. (2011), respectively, are also shown for comparison.}
 \label{4lfs}
\end{figure}

Figure~\ref{4lfs} shows the luminosity functions for the SDSS Petrosian and {\tt cmodel} magnitudes (corrected for the SDSS sky subtraction problems as described in Bernardi et al. 2010), and SerExp and Sersic magnitudes (from {\tt PyMorph}).  Although the difference between the Petrosian and {\tt cmodel} magnitudes has been known for some time, the fact that single-Sersic based counts lie substantially above those based on the SDSS outputs has only recently begun to attract attention.  For example, the GAMA based results of Hill et al. (2011) point to this difference, but because GAMA covers a substantially smaller volume than the SDSS, it does not probe the high luminosity end which is of most interest here.  Our {\tt PyMorph} reductions, which are in good agreement with Hill et al. at $M_r>-23$, show that at $M_r<-23$ the difference with respect to {\tt cmodel} counts is dramatic indeed.  

The {\tt PyMorph}-based counts are in good agreement with those which use the Simard et al. (2011) single-Sersic reductions, except at luminosities brighter than $M_r\sim -24$, where {\tt PyMorph} tends to be brighter (c.f. Figure~\ref{compfit}), so the {\tt PyMorph} luminosity function shows more high luminosity objects.  This agreement illustrates that our finding that single-Sersic fits return substantially more objects in the high luminosity tail than do {\tt cmodel} magnitudes is robust to changes in the reduction pipeline.  

The solid curves show the result of fitting 
\begin{equation}
 X\phi(X) = \phi_\alpha\beta\,\left(\frac{X}{X_*}\right)^\alpha\, 
                     \frac{{\rm e}^{-(X/X_*)^\beta}}{\Gamma(\alpha/\beta)}
        + \phi_\gamma\, \left(\frac{X}{X_\gamma}\right)^\gamma\,{\rm e}^{-(X/X_\gamma)}
 \label{phiX}
\end{equation}
with $X=L$ to the counts.  
The associated luminosity density is 
$\rho_X = \phi_\alpha\,X_*\,\Gamma[(1+\alpha)/\beta]/\Gamma[\alpha/\beta] + 
 \phi_\gamma\,X_\gamma\,\Gamma[1+\gamma]$.  
The first term in equation~(\ref{phiX}) is the same functional form as that used by Bernardi et al. (2010); the second is required to fit the slight bump at the faint-end.  The parameters which yield the best-fit are given in Table~\ref{fits}.  Note that the value of $X_*$ is not as intuitive as is its mean value 
$X_*\,\Gamma[(1+\alpha)/\beta]/\Gamma[\alpha/\beta]$.  

\begin{figure}
 \centering
 \includegraphics[scale = .4]{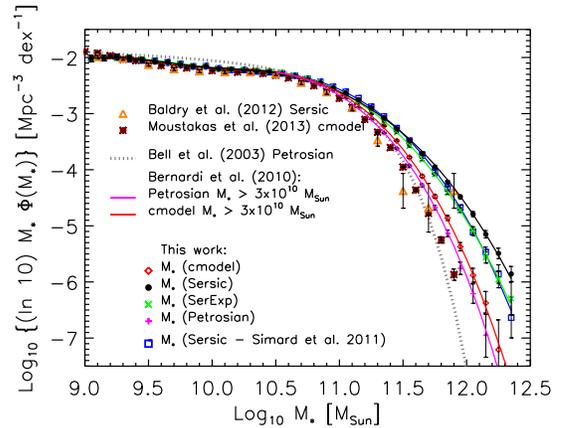}
 \caption{Same as previous figure, but now for the associated 
  stellar mass functions.  Recent stellar mass functions from 
  Baldry et al. (2012; based on Sersic magnitudes) and 
  Moustakas et al. (2013; based on {\tt cmodel} magnitudes) are also shown.  
  All stellar masses assume a Chabrier IMF.}
 \label{4mfs}
\end{figure}

The observed distributions shown here have been broadened slightly by measurement errors.  Bernardi et al. (2010) showed how to modify the analog of equation~(\ref{phiX}) so as to estimate the parameters of the intrinsic distribution, but that, in practice, the difference between the intrinsic and observed broadened distributions is small -- much smaller than the difference between the {\tt PyMorph} and {\tt cmodel} counts, so here we show the results of the observed distribution not the intrinsic one.  

\subsection{The stellar mass function}
Figure~\ref{4mfs} shows the associated stellar mass functions.  In all cases, $M_*$ was estimated from the luminosity and the {\tt cmodel} $g-r$ color assuming the Chabrier (2003) IMF as described in Bernardi et al. (2010).  We use the {\tt cmodel} color because the main goal of this paper is to study the effect on $\phi(M_*)$ from changes in $L$.  By using {\tt cmodel} colors, we are ensuring that our $M_*/L$ estimates for each object are the same as in Bernardi et al. (2010); however, the $L$ estimate for each object differs (Petrosian $\ne$ {\tt cmodel} $\ne$ {\tt PyMorph}).  Notice again that the {\tt PyMorph}-based estimates (as well as those from Simard et al.) lie well-above the {\tt Petrosian} and {\tt cmodel} ones, although some of the difference, especially with respect to {\tt Petrosian}, is due to sky-subtraction issues.  (Of course, if the stellar population models used to estimate $M_*/L$ are incorrect, or if the IMF is mass-dependent, then this will modify the results.  See Section~\ref{comparison} for comparison with other work.)

\begin{table*}
\caption[]{The binned $\phi(M_r)$ counts for the full sample and when weighted by the probability of a given morphological type. X = M$_r$ $[$mag$]$ and Y = Log$_{10} \phi$(M$_r$) $[$Mpc$^{-3}$ dex$^{-1}$$]$. Four electronic tables are provided in this format based on the type of magnitude: {\tt PyMorph} Sersic (LF-Ser.dat), {\tt PyMorph} SerExp (LF-SerExp.dat), {\tt Simard} Sersic (LF-Ser-Simard.dat) and cmodel from Bernardi et al. (2010) (LF-cmodel.dat).\\}
\begin{tabular}{cccccc}
 \hline 
  X & Y (All)  & Y wP(Ell) & Y wP(S0) & Y wP(Sab) & Y wP(Scd)  \\
 \hline
-17.700  &  -2.350$\pm$0.065 &  -4.030$\pm$0.065 &   -3.209$\pm$0.065 &  -2.708$\pm$0.065 &  -2.706$\pm$0.065\\
 \hline 
\end{tabular}
\label{tabL} 
\end{table*}

\begin{table*}
\caption[]{The stellar mass function $\phi(M_*)$ of the full sample, and when weighted by the probability of being a given morphological type. X = Log$_{10}$M$_*$ $[$M$_{\odot}$$]$ and Y=Log$_{10}$$[$(ln 10) M$_*$ $\phi($M$_*$)$]$ $[$Mpc$^{-3}$ dex$^{-1}$$]$. Four electronic tables are provided in this format based on the type of magnitude: {\tt PyMorph} Sersic (MsF-Ser.dat), {\tt PyMorph} SerExp (MsF-SerExp.dat), {\tt Simard} Sersic (MsF-Ser-Simard.dat) and cmodel from Bernardi et al. (2010) (MsF-cmodel.dat).\\}
\begin{tabular}{cccccc}
 \hline 
  X & Y (All)  & Y wP(Ell) & Y wP(S0) & Y wP(Sab) & Y wP(Scd) \\
 \hline
9.050 &  -2.012$\pm$0.053 &  -3.884$\pm$0.053 &  -2.933$\pm$0.053 &  -2.376$\pm$0.053  & -2.339$\pm$0.053\\
 \hline 
\end{tabular}
\label{tabMs} 
\end{table*}

\begin{figure}
 \centering
 \includegraphics[scale = .4]{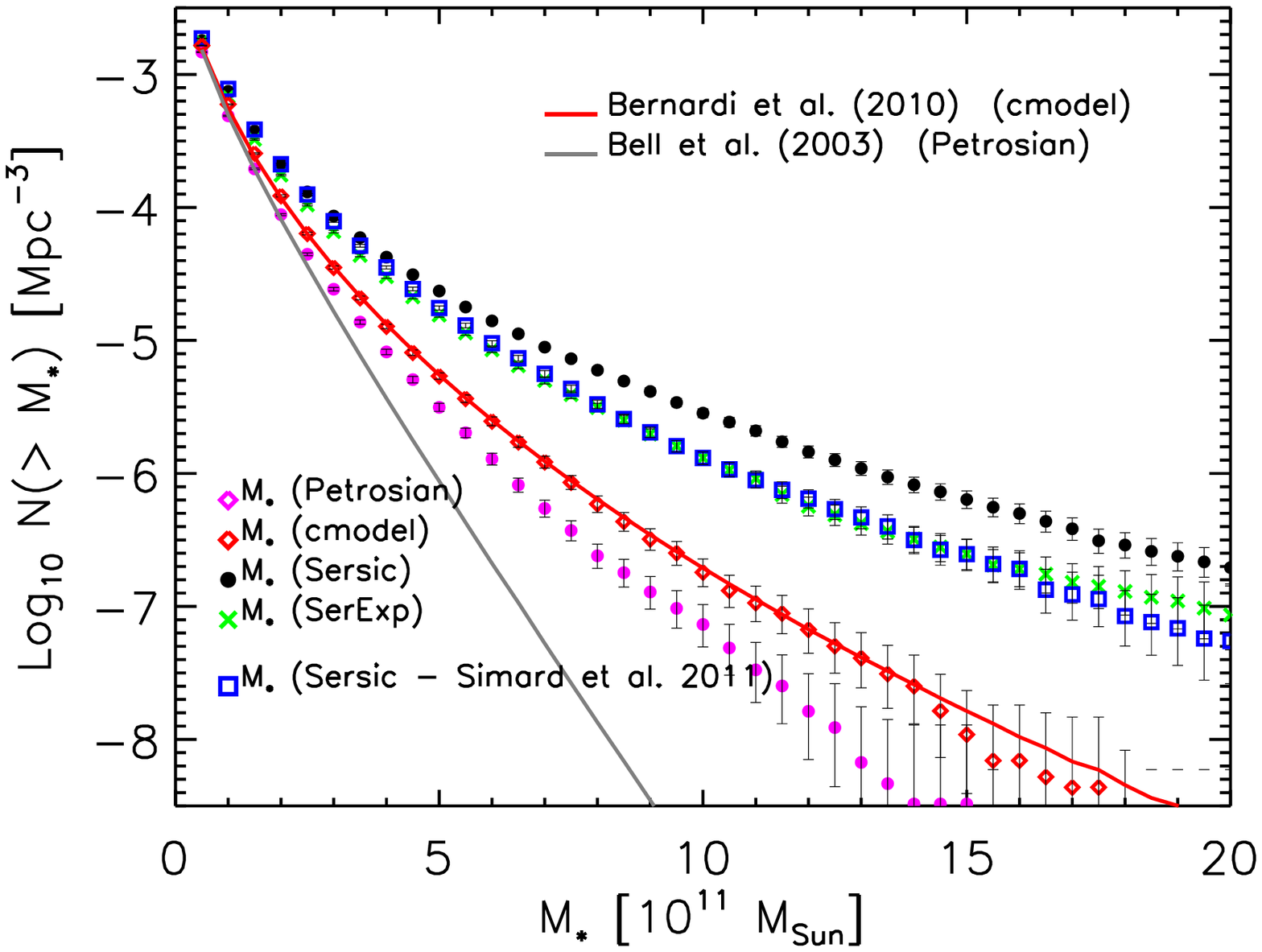}
 \includegraphics[scale = .4]{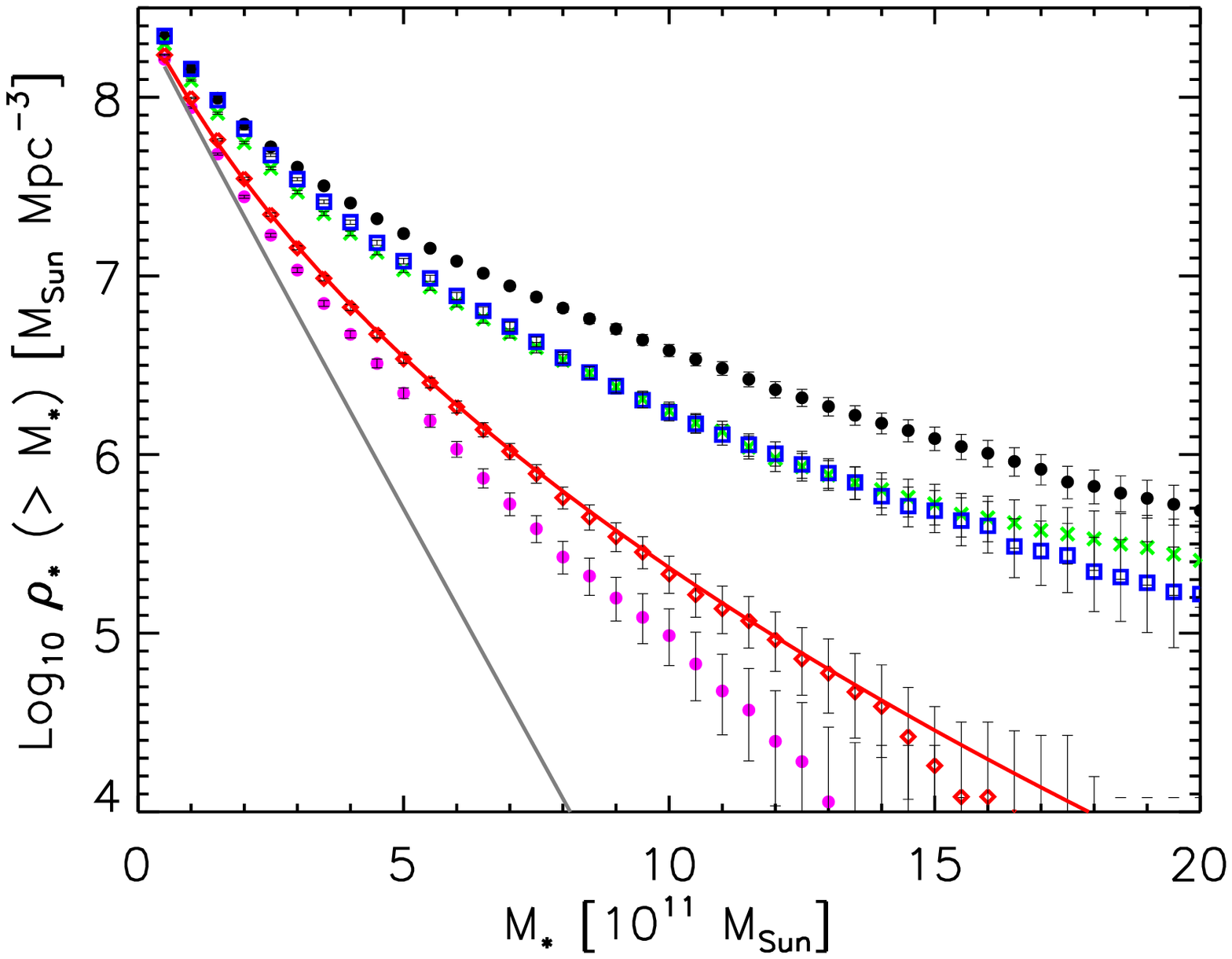}
 \caption{Similar to previous figure, but now showing cumulative 
  rather than differential counts. Top and bottom panels show number 
  and stellar mass density respectively.  To facilitate comparison with 
  previous work we show the fit of Bell et al. (2003).}
 \label{cumulative}
\end{figure}

The estimate from Baldry et al. (2012) lies below all the others.  This is remarkable because it is based on the GAMA-Sersic reductions, and we have already seen that the associated $\phi(L)$ (from Hill et al. 2011) is in good agreement with that based on {\tt PyMorph}.  Therefore, the difference in $\phi(M_*)$ must be entirely due to $M_*/L$, even though Baldry et al. also assume a Chabrier IMF.  We discuss this more in Section~\ref{comparison}.  
 
We think it is interesting to present our results in a format which highlights just how much the {\tt PyMorph}-based values differ from other work (we use Bell et al. 2003 for comparison).  Figure~\ref{cumulative} shows cumulative (rather than differential) counts, both for number and stellar mass-weighted density.  The number counts at the mass scale above which the number density of objects is $10^{-6}$Mpc$^{-3}$ is larger by a factor of $\sim 2$ compared to the {\tt cmodel}-based counts (a factor of $\sim 3$ compared to Bell et al.).  Alternatively, at $M_*=6\times 10^{11}M_\odot$, the {\tt PyMorph} counts lie a factor of $\sim 8$ above those based on {\tt cmodel} magnitudes (a much larger factor above Bell et al.).  For the mass-weighted counts the corresponding discrepancies at $10^{6}M_\odot$Mpc$^{-3}$ or $6\times 10^{11}M_\odot$ are similar or slightly larger.  

To make our results simple to use, in addition to Table~\ref{fits}, which reports the parameter values associated with the best-fits, we have made the binned counts available in the electronic versions of Tables~\ref{tabL} and~\ref{tabMs}.

\section{Comparison with previous work}\label{comparison}
Recently, it has become fashionable to concentrate more on $\phi(M_*)$ than $\phi(L)$.  Unfortunately, this combines two very different types of uncertainty:  that associated with the total light, and the other associated with the stellar mass-to-light ratio $M_*/L$ (e.g. mismatched stellar templates, mass-dependence of the IMF, etc.).  To illustrate this, we compare our results with two of the most recent determinations of $\phi(M_*)$:  those of Baldry et al. (2012) and Moustakas et al. (2013).  Although we all assume a Chabrier IMF, their estimates of $\phi(M_*)$ are more similar to one another than they are to ours. However, as we argue below, this implies large differences in their $M_*/L$ values since they used different luminosities to get $M_*$.  

The Baldry et al. analysis is based on single Sersic fits to the light profiles from galaxies with $z < 0.06$ in the GAMA survey reported by Kelvin et al. (2012).  
Of the $7335$ GAMA matches in the Bernardi et al. (2010) sample which we are studying here, only $1612$ have $z < 0.06$.  The red triangles in Figure~\ref{comp} show that, for these 1612 galaxies, the single Sersic {\tt PyMorph} estimates of the total light are in good agreement with those derived by Kelvin et al., and used by Baldry et al.  However, the redshift cut eliminates most of the high luminosity objects which are of most interest to our paper.  



\begin{figure}
 \centering
 \includegraphics[scale = .4]{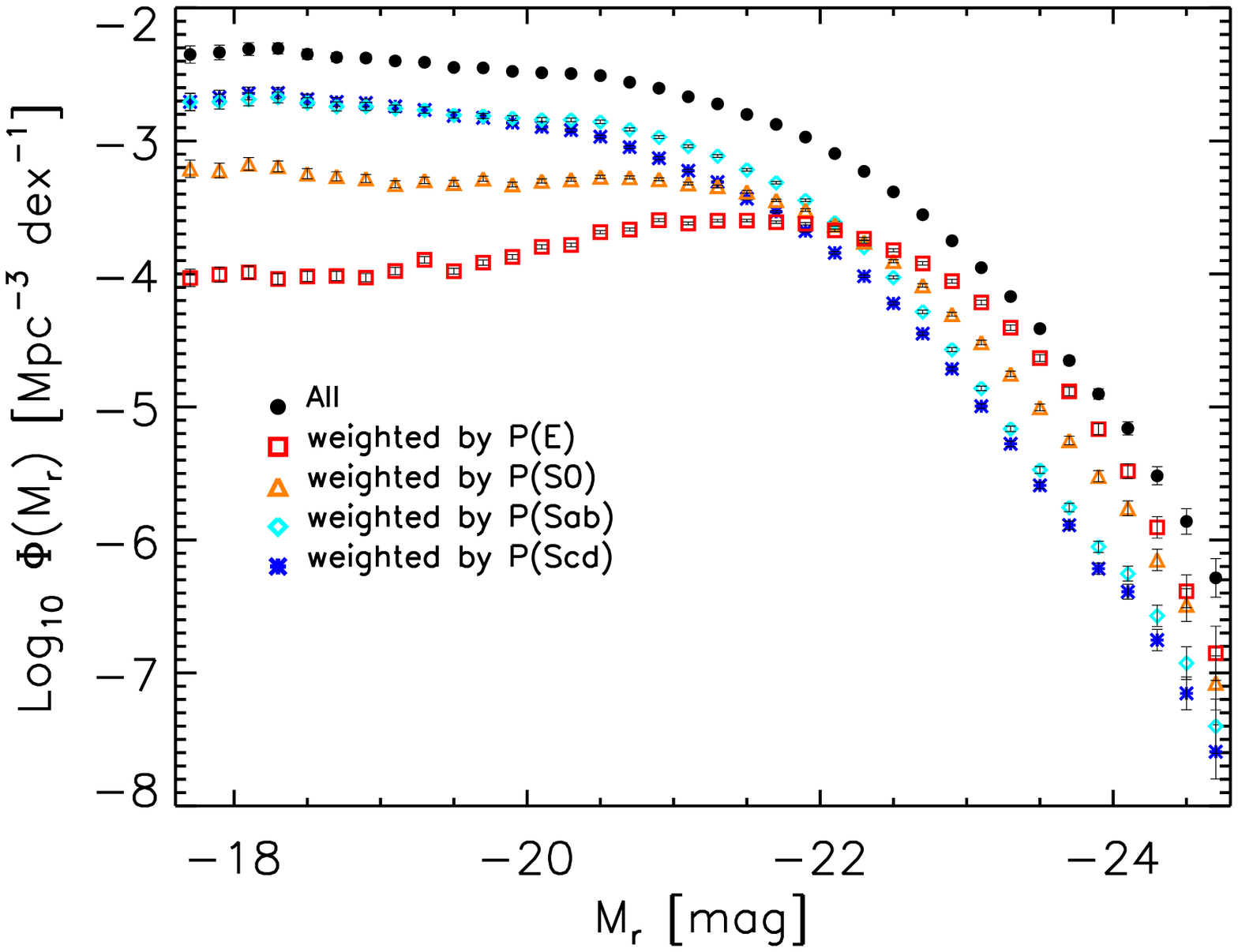}
 \includegraphics[scale = .4]{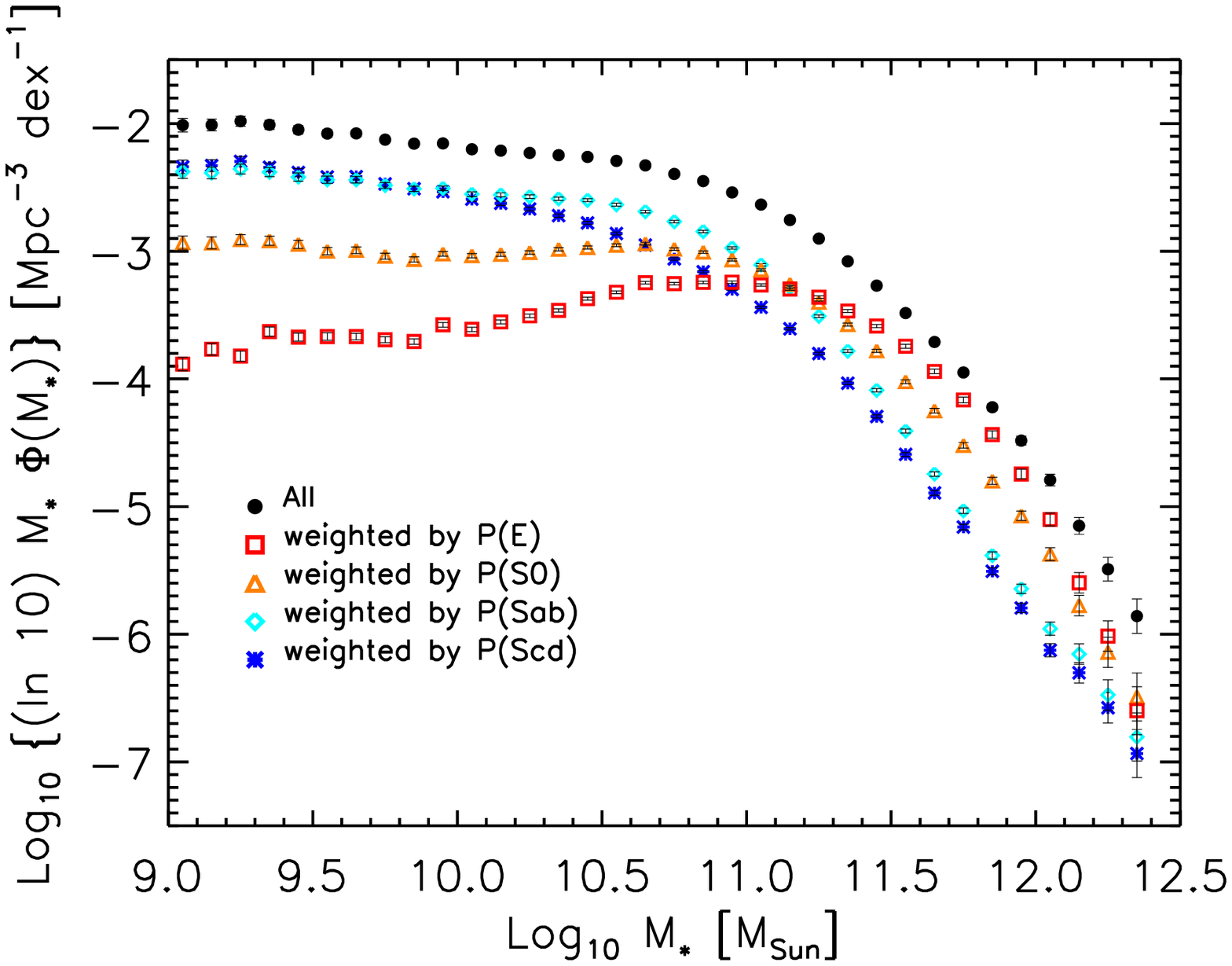}
 \caption{Morphological dependence of the luminosity (top) and stellar mass (bottom) functions.  In all cases, the luminosities are based on Sersic magnitudes, stellar masses assume a Chabrier IMF, and morphologies are based on the BAC method of Huertas-Company et al. (2011).  }
 \label{morphs}
\end{figure}

We have checked explicitly and found that the GAMA-based luminosity function for these $z<0.06$ objects is actually in good agreement with that based on {\tt PyMorph} single Sersic reductions when restricted to $z<0.06$, and this is in good agreement with that for the full GAMA sample of Hill et al. (2011) shown in Figure~\ref{4lfs}.  (Of course, the smaller volume means the error bars are large, and the comparison is effectively limited to abundances greater than about $10^{-4}$ Mpc$^{-3}$ dex$^{-1}$.)  However, our {\tt PyMorph}-based $\phi(M_*)$ estimate lies well above that of Baldry et al., so we conclude that our $M_*/L$ values must be larger than theirs.  The Baldry et al. estimate is much closer to, though slightly below, the estimate of Bell et al. (2003, scaled to a Chabrier IMF).  However, since the Baldry et al. Sersic-based $\phi(L)$ estimate agrees with ours, and this last lies well above the $\phi(L)$ associated with the SDSS Petrosian magnitudes which were used by Bell et al. (see our Figure~\ref{4lfs}) we conclude that the Baldry et al. $M_*/L$ values must also be smaller than those of Bell et al.  

Moustakas et al. use {\tt cmodel} magnitudes.  Since their sample is essentially the same SDSS sample as ours, we expect their luminosity function to agree with that of Bernardi et al. (2010).  However, Figure~\ref{4mfs} shows that although their stellar mass function is similar to our {\tt cmodel}-derived $\phi(M_*)$ at low masses, it is different at the high end, indicating that the Moustakas et al. $M_*/L$ ratios are smaller than ours.  (We suspect that this must be related to the choice of template used to estimate $M_*/L$ at higher masses, for reasons given in Fig.~22 and associated discussion of Bernardi et al. 2010.)  On the other hand, despite differences at lower masses, the Moustakas et al. $\phi(M_*)$ is reasonably well approximated by the Bell et al. estimate at higher masses.  Since the {\tt cmodel} luminosity function is different from the Petrosian one at these high masses, we conclude that the Moustakas et al. $M_*/L$ values must be smaller than those of Bell et al., and different again from those of Baldry et al. (who used Sersic rather than {\tt cmodel} magnitudes).  

Therefore, 
\begin{itemize}
 \item by comparing the {\tt PyMorph} Sersic-based $\phi(M_*)$ with that based on Simard et al. our Figure~\ref{4mfs} quantifies how differences due to the uncertainties in a given light profile fit (mainly due to sky subtraction issues, see Meert et al. 2013b) affect $\phi(M_*)$; 
 \item by comparing our {\tt SerExp} and {\tt Sersic}-based $\phi(M_*)$ fits, our Figure~\ref{4mfs} quantifies the effect of fitting different light profile models;
 \item by comparing our {\tt Petrosian} based $\phi(M_*)$ to the Bell et al. fit, our {\tt cmodel}-based estimate to that from Moustakas et al., or our Sersic (or Simard et al.) based $\phi(M_*)$ to Baldry et al, our Figure~\ref{4mfs} quantifies how systematic differences in $M_*/L$ affect $\phi(M_*)$.
\end{itemize}
This shows that the effects on $\phi(M_*)$ of using the total luminosity computed from different fits to the light profile are dramatic; it is important to specify how the light profile was fit when reporting a luminosity or stellar mass function.  

\section{Dependence on morphology}\label{morph}
We have combined our {\tt PyMorph} SerExp reductions with the Bayesian Automated morphological Classifier of Huertas-Company et al. (2011) to determine how the luminosity and stellar mass functions depend on morphology.  This algorithm returns the probability that an object is one of four types (for each object, the sum of the four probabilities is unity).  Therefore, to estimate the luminosity function, we simply weighted object $j$ by $p_j({\rm type})/V_{\rm max}(L^{\rm Pet}_j)$.  

In practice, there are a number of faint objects for which $p_j({\rm E})\le 0.15$.  These can dominate over the counts from similarly faint objects for which $p_j({\rm E}) > 0.85$.  If these low values of $p$ are simply the result of errors in the BAC algorithm, then these will wrongly boost the luminosity function at the faint end.  To check the magnitude of the effect, we have set to zero all values of $p\le 0.15$, and reassigned the weight to the types which remain with contribution proportional to the nonzero remaining values of $p$ such that the sum over the four $p$ values (some of which are now zero) is still unity.  This reduces the counts of faint ellipticals and luminous spirals by a factor of about two:  it is these modified counts which we show in Figure~\ref{morphs}.  

The binned counts are given in Tables~\ref{tabL} and ~\ref{tabMs}, which are provided in convenient electronic format in the online version of the journal.  (Note that summing up the luminosity functions of the different morphological types does not quite give the luminosity function of the full sample.  The small discrepancy arises because we were unable to find BAC classifications for a small fraction of the objects in our sample.)  Figure~\ref{morphs} confirms the well-known trend for early-types (E and S0) to dominate the high-mass end, and later-types (Sa to Sd) to dominate at lower masses.

We have compared these estimates with those of Nair \& Abraham (2010), who used the T-Type classification using the modified RC3 (Third Reference Catalogue of Bright Galaxies) classifiers.  We set E (T$=-5$ and T$=-4$), S0 (T$= -3$, T$=-2$ and T$=-1$), Sa (T$= 0$, T$=1$ and T$=2$), Sb (T$=3$ and T$=4$), and Scd (T$= 5$, T$=6$ and T$=7$).  Our E and S0 counts are in quite good agreement, though, at the bright end, they tend to classify objects as E rather than S0 when BAC divides its weights between E and S0.  Differences are slightly more pronounced for later types:  our Scd counts are in good agreement at the faint end, but lie substantially above theirs at high luminosities (where the counts are falling exponentially).  On the other hand, our Sab counts lie about a factor of two below theirs at high $L$, suggesting that BAC assigns some weight to the extreme Scd classification when Nair et al. choose the extremes.

\section{Discussion}
{\tt PyMorph} Sersic or Sersic+exponential based estimates of the total light of a galaxy are larger than those based on SDSS pipeline outputs (Petrosian or {\tt cmodel}; Figure~\ref{compfit}).  As a result our {\tt PyMorph}-based luminosity and stellar mass functions are rather different from previous work:  they have more light and mass at the bright, massive end (Figures~\ref{4lfs} and~\ref{4mfs}). 

Petrosian magnitudes have been popular in recent years.  However, our SDSS Petrosian-based luminosity functions, which are similar to those in the literature, lie well-below any of the others.  This is consistent with a number of recent studies which agree that they, and {\tt cmodel} magnitudes, underestimate the total light (Simard et al. 2011; Hill et al. 2011; Kelvin et al. 2012; Bernardi et al. 2013; Meert et al. 2013b; Hall et al. 2012).  Some of this is due to sky-subtraction problems which affect the SDSS pipelines (Bernardi et al. 2010) more than {\tt PyMorph} (Meert et al. 2013b).  While Petrosian magnitudes with better sky-subtraction are brighter (e.g. He et al. 2013, which appeared after the initial submission of our work), thus reducing the difference with respect to our {\tt PyMorph} magnitudes, our results suggest that they should only be used after applying the sorts of corrections advocated by Graham et al. (2005) which partially account for the fact that Petrosian magnitudes underestimate the total light when the light profile has extended wings (e.g. Blanton et al. 2001; Trujillo et al. 2001; Andreon 2002).  For the brightest galaxies, even such corrected Petrosian magnitudes are fainter than our {\tt PyMorph} magnitudes.  In addition, Petrosian-derived quantities are not seeing corrected; this impacts studies of scaling relations (e.g. see Appendix~A in Hyde \& Bernardi 2009) more than our current study of the mass function.  For these reasons, we believe that fits to {\tt Sersic} or {\tt SerExp} profiles are to be preferred, especially in higher redshift datasets where seeing is an issue.  


Our Sersic-based luminosity functions, from both {\tt PyMorph} and Simard et al. (2011), are very different from the Sersic-based analysis in Blanton et al. (2003).  However, the Sersic parameters used by Blanton et al. were estimated from a 1-dimensional radial surface brightness profile, measured in $\sim 5-10$ azimuthally averaged annuli.  This procedure is expected to be significantly less accurate than the 2-dimensional fits to the whole galaxy image performed by {\tt PyMorph} and Simard et al.  In addition, these more recent analyses include a more careful treatment of the background sky, especially in crowded fields, so we believe that these more recent Sersic reductions (as well as those of Kelvin et al. 2012 for the subset of objects in the GAMA survey) supercede those of Blanton et al.  


Our fits indicate that the luminosity density at $z\sim 0.1$ is about 10\% larger than our previous work with {\tt cmodel} magnitudes (Table~\ref{fits}), and about a factor of two larger than when based on Petrosian SDSS DR7 magnitudes (Figure~\ref{cumulative}).  This difference is driven by the most luminous objects which are predominantly quiescent early-type galaxies (Figure~\ref{morphs}).  Since a number of authors now agree that SDSS pipeline magnitudes underestimate the true luminosity, and these more recent algorithms are in reasonably good agreement with the {\tt PyMorph} reductions used by us (Figure~\ref{comp}), we conclude that there is now good agreement that the bright end of the luminosity function may be substantially brighter than the SDSS DR7 Petrosian magnitudes suggest. 

As one might expect given our analysis of the luminosity function, our (Chabrier IMF-based) stellar mass densities at $z\sim 0.1$ are about 25\% larger than our previous work (Bernardi et al. 2010) with {\tt cmodel} magnitudes (Table~\ref{fits}), which was itself considerably larger than when based on the SDSS DR7 Petrosian magnitudes which are often used for this purpose (Figure~\ref{cumulative}).  As a result, our estimates have implications for studies of the evolution of the star formation rate, the growth of stellar mass in galaxies, the processes by which this mass was assembled, and Halo Model analyses of the $M_*-M_{\rm halo}$ relation.  For example:  
\begin{itemize}
\item Our higher stellar mass density at $z\sim 0$ resolves the tension with respect to the total mass density inferred from the integrated star formation rate (SFR), as noted in Bernardi et al. (2010).
\item A higher number density of massive galaxies in the local Universe allows for a higher incidence of major (in addition to minor) mergers in driving the stellar mass growth of the most massive central galaxies at late times (e.g., Bernardi et al. 2011a,b; Shankar et al. 2013).  This conclusion rests, of course, on the quality of the determination of the high redshift stellar mass function, a task which we expect to be even more challenging than in the local Universe.  In this respect, it is interesting that the $z\sim 1$ counts at $M_*\ge 10^{11.5}M_\odot$ of Carollo et al. (2013) are in rather good agreement with our $z\sim 0$ (Sersic-based) estimate, strongly limiting merger rates at the high mass end.  
\item A higher number density at high masses would better match the stellar and dynamical mass functions, possibly reducing the need for a strong mass-dependent variation of the IMF (see, e.g., Fig. 23 in Bernardi et al. 2010), although the true extent of the latter statement relies on accurate dynamical mass measurements with appropriate effective radiuses and structure constants.  We plan to address this separately.  
\item The stellar mass function in the local Universe is one of the fundamental ingredients in popular semi-empirical models for populating haloes with galaxies, such as the halo occupation and abundance matching techniques (e.g., Cooray \& Sheth 2002; Berlind \& Weinberg 2002; Vale \& Ostriker 2004; Shankar et al. 2006; Zehavi et al. 2011; Leauthaud et al. 2012; Moster et al. 2013).  If the more massive galaxies are more abundant, then Halo Model analyses will assign them to lower mass halos.  Since lower mass halos are less strongly clustered, we expect the most massive galaxies to be less strongly clustered than current models assume.  
\item This would also imply that the median baryon fraction at the high mass end may be significantly higher than previously thought. This may pose serious questions about the impact of feedback from active galactic nuclei, in the
quasar and/or radio modes (e.g., Granato et al. 2004; Croton et
al. 2006; Silk \& Mamon 2012); even more so, if one considers that the baryon 
fraction of ellipticals at their formation epoch ($z>1$) must have been even higher.
\item More stellar mass at the high mass end directly impacts studies of how the stellar fraction compares with that in the gas detected by X-ray and Sunyaev-Zeldovich experiments.  E.g., by decreasing the halo mass one should associate with a given stellar mass, it potentially reduces the discrepancy shown in Figure~9 of Ade et al. (2013).  
\item These effects on the stellar to halo mass mapping go in the same direction as those from plausible changes to the IMF at the massive end (Bernardi et al. 2010 and references therein), so the data may be indicating that a major revision of the results on the galaxy-halo mapping at the high mass end, is called-for.
\end{itemize}
Since a number of groups are currently engaged in such studies, we have provided the luminosity and stellar mass functions shown in Figures~\ref{4lfs}, \ref{4mfs} and \ref{morphs} in a convenient electronic tabular format in the online version of the journal (see Tables~\ref{tabL} and~\ref{tabMs}).  

We caution that our estimates of the amount of stellar mass in the most massive objects are at least $2\times$ larger than other recent determinations (e.g. Baldry et al. 2012 and Moustakas et al. 2013), even though we argued that our estimates of the luminosity function are in much better agreement than this $\phi(M_*)$-based number suggests.  Section~\ref{comparison} shows that their $M_*$ values are similar only because they assume very different $L$ and $M_*/L$ values from one another.  Therefore, until the field converges on what it believes to be reliable $M_*/L$ estimates for the highest mass objects (see Mitchell et al. 2013 for ongoing discussion of this point), we believe our results argue against calibrating models to published stellar mass functions:  calibrating to the luminosity function is more robust, as it does not combine differences in $L$ and $M_*/L$ into a single number $M_*$.  

When using our results, it is important to bear in mind that the {\tt PyMorph} estimates assume that the galaxy is either a single Sersic, or a two-component Sersic+Exponential system.  These both allow for substantial light beyond the core of the image, so there is some question as to just what it is that the profiles are fitting.  The most massive objects tend to be the brightest cluster galaxies (BCGs), so some of the excess light returned by Sersic and/or SerExp fits may contain intracluster light (ICL).  If so, then our estimates of the total stellar mass are appropriate for models which associate the ICL with the central galaxy.  This may even be physically reasonable, since most of the accretion and stripping which occured as the cluster assembled likely happened during accretion onto what is now the central object.

\section*{Acknowledgements}
We are grateful to L. Simard for sharing information about sky levels, 
S. Andreon, S. Courteau and A. Graham for helpful comments about 
previous work, and A. Kravtsov for urging us to complete this work 
which is supported in part by ADP/NNX09AD02G and NSF-AST 0908241.

\end{document}